\documentclass[aps,pra,twocolumn,showpacs,groupedaddress]{revtex4-1}   
\usepackage{hyperref}  
\usepackage{graphicx}  
\usepackage{dcolumn}   
\usepackage{bm}        
\usepackage{amssymb}   
\usepackage{verbatim}  
\usepackage{epstopdf}
\usepackage{color}
\newcommand{\ket}[1]{\ensuremath{\left| #1 \right\rangle}}
\newcommand{\bra}[1]{\ensuremath{\left\langle #1 \right|}}
\newcommand{\inner}[2]{{\langle{#1}|{#2}\rangle}}

\newcommand{\mele}[3]{\ensuremath{\left\langle #1 \right|#2\left| #3
\right\rangle}}
\newcommand{\dyad}[2]{{\ket{#1}\!\!\bra{#2}}}

\newcommand{\tr}{{\rm Tr}}

\newcommand{\beq}{\begin{equation}}
\newcommand{\eeq}{\end{equation}}
\newcommand{\bea}{\begin{eqnarray}}
\newcommand{\eea}{\end{eqnarray}}

\newcommand{\eq}[1]{{(\ref{#1})}}
\newcommand{\commentout}[1]{{}}

\newcommand{\half}{{\hbox{$\frac{1}{2}$}}}

\definecolor{red}{rgb}{1,0,0}

\begin{document}

\title{Optimal measurement precision of a nonlinear interferometer}

\author{Juha Javanainen}
\affiliation{Department of Physics, University of Connecticut, Storrs, Connecticut 06269-3046}

\author{Han Chen}
\affiliation{Department of Physics, University of Connecticut, Storrs, Connecticut 06269-3046}

\begin{abstract}
We study the best attainable measurement precision when a double-well trap with bosons inside acts as an interferometer to measure the energy difference of the atoms on the two sides of the trap. We introduce time independent perturbation theory as the main tool in both analytical arguments and numerical computations. Nonlinearity from atom-atom interactions will not indirectly allow the interferometer to beat the Heisenberg limit, but in many regimes of the operation the Heisenberg limit scaling of measurement precision is preserved in spite of added tunneling of the atoms and atom-atom interactions,  often even with the optimal prefactor.
\end{abstract}
\maketitle


\section{Introduction}
Paraphrased for optical or atomic interferometry, the Heisenberg limit~\cite{HOL93} states that the best possible achievable uncertainty of phase measurements is inversely proportional to the number of bosons $N$. The Heisenberg limit, with continuous refinement and sharpening of the concept~\cite{GIO06,ZWI10}, is one of the enduring paradigms in quantum metrology. Nevertheless, there is nothing sacred about $1/N$ scaling; in nonlinear schemes where the quantity to be measured couples to a  $k$-body operator the measurement uncertainty could scale as $1/N^k$~\cite{LUI04,BOI07}. This is well and fine if the aim is the best possible measurement of, say, the atom-atom scattering length~\cite{REY07,CHO08,TAC10}, which in fact enters the time evolution in conjunction with a two-body operator. However, such  a $1/N^2$ scaling does not directly help if you insist on a measurement of a quantity that {\em does\/} couple to the first power of boson number. The nonlinear (in light intensity) Faraday rotation, a possible method to measure a magnetic field, demonstrably allows for improved noise properties that originate from the nonlinearity~\cite{NAP11}, but virtually all of practical interferometry is based on one-particle coupling.

The impetus to the present work is the question of what happens to the Heisenberg limit, prefactor or scaling, when the time evolution of a probe depends on the quantity to be measured and also has other components that modify the evolution, particularly nonlinearity. Our analysis takes place in two steps.

In Sec.~\ref{GENMEAS} we address the general case. Schematically, the dynamics that converts the quantity to be measured $\theta$ into a state change of the probe has two parts, one described by a Hamiltonian $\theta G$, where  $G$ could be called generator of $\theta$ translations, and an additive part $\bar K$ that does not depend on $\theta$. We show how time independent perturbation theory may be harnessed to analyze the best possible precision of the measurements of the quantity $\theta$, and subsequently argue that the addition of the term $\bar K$, {\em any\/} term $\bar K$, cannot improve the precision~\cite{GIO06,BOI07}.

Next we formulate a tangible example, measurement of the energy difference of an atom in two potential wells in the presence of both tunneling of the atoms between the wells and atom-atom interactions. The same physical system was studied for similar aims in Ref.~\cite{GRO11}, although from a more practice oriented standpoint.  The optical analog would be a device that allows tunneling (exchange) of photons between the two arms of the interferometer and also has nonlinear phase shifts, as per the Kerr effect.

We state the example in Sec.~\ref{MEASMOD}, and analyze the achievable limit of precisions in Sec.~\ref{MEASDEP} as a function of the parameters of the system. Some results, such as how the interplay of tunneling and atom-atom interactions affect the precision, should be educational per se, but more to our  point of principle, we have here explicit examples in which the pieces added to the time evolution of the probe reduce the best possible attainable measurement precision. For the most part the Heisenberg limit $1/N$ scaling is still retained, but with a reduced prefactor. A few remarks in Sec.~\ref{CONCLUSIONS} wrap up the paper.

\section{Generic measurement scheme}\label{GENMEAS}
We start with the general scheme of quantum measurements: Prepare an initial state for the ``probe'', have the probe evolve under some quantum mechanical law that depends on the quantity to be measured, and finally infer the value of the quantity from measurements on the probe. The best possible precision in the setup we are considering is achieved with a pure initial state~\cite{FUJ01}, so we take the measurement to start with a pure state $\ket{\psi_0}$. Moreover, we assume that the quantum mechanical evolution is generated by a hermitian Hamiltonian. A pure state then remains pure during the evolution. We have a unitary mapping of the initial state $\ket{\psi_0}$ to the state $\ket{\psi(\theta)}$ that depends on the parameter $\theta$ to be measured, and we may write
\beq
\ket{\psi}\equiv\ket{\psi(\theta)} = e^{-i K(\theta)} \ket{\psi_0}\,,
\label{DEFTRAN}
\eeq
where $K(\theta)$ is Hermitian. Let us also define 
\beq
\ket{\psi'}\equiv\ket{\psi'(\theta)} \equiv \frac{d}{d\theta} \ket{\psi(\theta)}.
\label{DEFDER}
\eeq

The criterion we use for the precision of the measurement is the variance, the square of the standard deviation. A well-defined procedure exists to find the smallest value of the measurement uncertainty over all possible measurements, given the initial state and the actual value of the parameter $\theta$ (!), and it even describes a measurement that could be carried out  (in principle) to reach the minimum uncertainty~\cite{HEL76,BRA94,BRA96}.  The variance is expressed as the inverse of quantum Fisher information $F$, which in turn is found from the symmetric logarithmic derivative of the density operator $\Lambda$. A straightforward calculation~\cite{BRA96} using the norm conservation
\beq
\inner{\psi}{\psi'} + \inner{\psi'}{\psi} = 0
\eeq
shows that the symmetric logarithmic derivate, the  Fisher information, and the standard deviation of the measurement results equal
\bea
\Lambda &=& 2 (\ket{\psi'}\bra{\psi} + \ket{\psi}\bra{\psi'}),\\
F &=& \tr(\rho \Lambda^2) = 4 [\inner{\psi'}{\psi'}-|\inner{\psi}{\psi'}|^2],
\label{FISHER}\\
\sigma_\theta &=& \frac{1}{\sqrt{F}}\,.
\eea

It pays to notice that everything we say in the present Sec.~\ref{GENMEAS} is completely general. Every measurement that starts from a pure state and is based on Hamiltonian time evolution where the quantity to be measured is a parameter in the Hamiltonian is covered.

\subsection{Perturbation theory}
The technical innovation here is to employ time independent perturbation theory in both numerical and analytical work. We explain presently.

The equality
\beq
\ket{\psi'} = \left( \frac{d}{d\theta} e^{-iK(\theta)}\right)\ket{\psi_0}
\label{OPDER}
\eeq
guides us to examine the derivative of the evolution operator $e^{-iK(\theta)}$ with respect to the parameter $\theta$. For an infinitesimally small $d\theta$ we have
\bea
d\theta\frac{d}{d\theta}e^{-i K(\theta)} &\simeq& e^{-i K(\theta +d\theta)} - e^{-i K(\theta)}\nonumber\\
&\simeq& e^{-i (K + d\theta\,K')}-e^{-i K}\,,
\label{EVDER}
\eea
where we have dropped a few explicit arguments $\theta$ and  defined
\beq
K' \equiv K'(\theta)  \equiv \frac{d}{d\theta} K(\theta)\,.
\eeq

Suppose we know the eigenvalues $\lambda_n$ and eigenvectors  $\ket n$ of $K(\theta)$, so that we have the spectral representation
\beq
K = \sum_n\lambda_n\dyad{n}{n}\,.
\eeq
One way to proceed with Eq.~\eq{EVDER} is to try and find the similar spectral representation for the operator $K + d\theta \,K'$ for an asymptotically small $d\theta$, eigenvalues and eigenvectors of the operator $K$ perturbed by the ``small'' operator $d\theta\, K'$. This is evidently an exercise in the usual time independent perturbation theory.

Let us first assume that the eigenvalues $\lambda_n$ of the operator $K$ are nondegenerate. To the leading nontrivial order in $d\theta$ the eigenvalues and eigenvectors of $K+d\theta\,K'$ are then of the form
\beq
\Lambda_n = \lambda_n +d\theta\,\xi_n,\quad\ket{\psi_n} = \ket n + d\theta\, \ket{\phi_n}\,,
\label{LAMBDAS}
\eeq
with
\beq
\xi_n = \mele{n}{K'}{n},\quad \ket{\phi_n} = \sum_{m\ne n} \ket m \,\frac{\mele{m}{K'}{n}}{\lambda_n-\lambda_m}\,.
\label{PERDEFS}
\eeq
If there are degeneracies, we are dealing with a combination of degenerate and nondegenerate perturbation theory.  We then choose the eigenstates of $K$ so that in any degenerate manifold they are also eigenstates of $K'$, and simply drop the seemingly divergent terms in the sum in~\eq{PERDEFS}. To the leading nontrivial order in $d\theta$ we then have
\bea
&&e^{-i (K + d\theta\,K')} =\sum_n e^{-i \Lambda_n} \dyad{\psi_n}{\psi_n}
 \nonumber\\&&\simeq\sum_n e^{-i\lambda_n} \dyad{n}{n}\nonumber\\
 &&+ d\theta\sum_n e^{-i\lambda_n}\left(
\dyad{\phi_n}{n} - i\xi_n \dyad{n}{n}+\dyad{n}{\phi_n}
\right)\,,
\eea
which gives 
\beq
\frac{d}{d\theta}e^{-i K(\theta)}  =\sum_n e^{-i\lambda_n}\left(
\dyad{\phi_n}{n} - i\xi_n\dyad{n}{n} +\dyad{n}{\phi_n}\right)\,.
\label{OPDEREXP}
\eeq

The model  that thoroughly permeates quantum metrology literature stipulates
\beq
K(\theta)= \theta G\,,
\label{SIMPLEMOD}
\eeq
where the hermitian $G$ could be called generator of $\theta$ translations. In this case we find
\beq
\ket{\psi'} = - iG \ket{\psi} = - iG e^{-i\theta G}\ket{\psi_0}\,,
\label{SIMPLEPSIP}
\eeq
both directly from Eq.~\eq{DEFDER}, and also indirectly from perturbation theory using Eqs.~\eq{OPDER}, \eq{OPDEREXP}, and~\eq{PERDEFS}. The Fisher information is then nothing but four times the variance of the generator $G$ in the state $\ket{\psi_0}$~\cite{HEL76,BRA94,BRA96}.

 However, in all but the simplest examples like this,  one would have to carry out the analysis numerically. Our method is to find the quantum Fisher information by combining Eqs.~\eq{FISHER}, ~\eq{OPDER},  \eq{OPDEREXP}, and~\eq{PERDEFS}. In numerical computations perturbation theory circumvents the need to take any derivatives numerically. A combination of non-degenerate and degenerate perturbation theory would make the computations tedious, so  in numerical analysis we ordinarily eliminate the degeneracies by adding a tiny perturbation to $K$ to break the symmetry that causes the degeneracies.

\subsection{Limit of precision}

We add one more layer of optimization and  find the initial state~$\ket{\psi_0}$ that produces the smallest possible measurement uncertainty, i.e., maximum Fisher information. In other words, we maximize $F(\ket{\psi_0})$ with respect to the initial state $\ket{\psi_0}$. There are special cases when this can be done analytically, too. For instance, in the model~\eq{SIMPLEMOD} the maximal Fisher information and a corresponding initial state are
\beq
F_M = (g_M-g_m)^2,\quad \ket{\psi_0} =\hbox{$ \frac{1}{\sqrt 2}$}(\ket{g_M}+e^{i\varphi}\ket{g_m})\,,
\label{SIMPLERES}
\eeq
where $g_M$ and $g_m$ are the largest and the smallest eigenvalue of the generator $G$, and $\varphi$ is an arbitrary relative phase between the respective eigenstates.

As general considerations go, we first write the derivative state from Eqs.~\eq{PERDEFS}, \eq{DEFTRAN}, and~\eq{OPDEREXP} in the form
\bea
\ket{\psi'} &=& \left(\frac{d}{d\theta}e^{-i K(\theta)}\right)\ket{\psi_0}= \left(\frac{d}{d\theta}e^{-i K(\theta)}\right) e^{i K(\theta)} \ket{\psi}
\nonumber\\
&=& -i L \ket\psi\,,
\label{COMPLEXPSIP}
\eea
where we have defined
\beq
L=\! \sum_{m,n} \frac{1\!-\!e^{i(\lambda_n\!-\!\lambda_m)}}{-i(\lambda_n-\lambda_m)}\ket{m}\mele{m}{K'}{n}\bra{n}.\label{EXPLFRM}
\eeq
In the formally singular term with $m=n$ the ratio is interpreted to have the value as appropriate for the limit $\lambda_n\rightarrow\lambda_m$, namely 1. The hermitian operator $L$ could be called local or instantaneous generator of $\theta$ translations at the given value of $\theta$.

As far as it comes to the largest attainable value of the Fisher information, we may just as well optimize with respect to the state $\ket\psi$,  the image of the initial state $\ket{\psi_0}$ in a norm preserving and one-to-one unitary mapping $e^{-iK}$. A comparison of Eqs.~\eq{SIMPLEPSIP}, \eq{SIMPLERES}, and~\eq{COMPLEXPSIP} immediately shows that the maximal Fisher information is
\beq
F_M = (\ell_M-\ell_m)^2\,,
\label{MFENT}
\eeq
where $\ell_M$ and $\ell_m$ are the largest and the smallest eigenvalue of the operator $L$.  An optimal input state is obtained by inverting the transformation $e^{-iK(\theta)}$, or
\beq
\ket{\psi_0} = \hbox{$\frac{1}{\sqrt2}$}\, e^{iK(\theta)}\,(\ket{\ell_M} + e^{i\varphi} \ket{\ell_m}\,.
\eeq

\subsection{Effect of added couplings on precision}
Suppose we start with the completely solvable model~\eq{SIMPLEMOD}, and ask how the measurement precision is affected if additional terms need to be considered in the dynamics of the probe. We write
\beq
K = \theta G + \bar K,
\eeq
where $\bar K$ stands for the added dynamics that is independent of the parameter $\theta$, so we have $K'=G$. Conversely, locally, around a given value of the parameter $\theta_0$, an arbitrary (differentiable) $K(\theta)$ can always be written in this form by choosing $G =  K'(\theta_0)$ and $\bar K = K(\theta_0)-\theta_0 G$.

From Eq.~\eq{EXPLFRM} the local generator $L$ may be written
\bea
L &=& \int_0^1 dx\, \sum_{m,n}\ket{m}e^{-ix\lambda_m}\mele{m}{G}{n}e^{ix\lambda_n}\bra{n}\\
&=& \int_0^1dx\,L(x)\,,\label{FINALRES}
\eea
with the definitions
\beq
L(x) = U(x)GU^\dagger(x),\quad U(x) = \sum_m\ket{m}e^{-ix\lambda_m}\bra{m}\,.
\eeq
$U(x)$ is unitary, so the spectra of all $L(x)$ are the same as the spectrum of $G$. But now, Eq.~\eq{FINALRES} can be viewed as a linear combination of a large number of operators $L(x)$ with positive coefficients that sum up to one. It is then easy to see from the variational principle that the largest eigenvalue $\ell_M$ of $L$ can be at most as large as the largest eigenvalue among the operators $L(x)$, i.e., the largest eigenvalue $g_M$ of $G$. Specifically, let $\ket{\ell_M}$ be a normalized eigenvector belonging to the largest eigenvalue of $L$, then we have
\bea
\ell_M &=& \mele{\ell_M}{L}{\ell_M} = \int_0^1 dx\, \mele{\ell_M}{L(x)}{\ell_M}\nonumber\\
&\le&  \int_0^1 dx\, g_M = g_M\,.
\eea
For the smallest eigenvalue we have similarly $g_m\le\ell_m$.

Therefore the maximal Fisher information $(\ell_M-\ell_m)^2$ in the amended measurement governed by the local generator $L$ is at most equal to the maximal Fisher information $(g_M-g_m)^2$ for the original generator $G$.  The added evolution $\bar K$ cannot improve the best attainable measurement precision. On the other hand, as we will see below, the range of the eigenvalues of $L$ may be narrower than the range of the eigenvalues of $G$, so that the attainable measurement precision may decrease.

Other authors give similar results from more~\cite{BOI07} or less~\cite{GIO06} related  arguments. Cast in terms of our present development, Ref.~\cite{GIO06} in fact claims that the attainable measurement precision remains unchanged. The error, again paraphrased for the present rendition of the mathematics, is in the implicit assumption that all of the operators $L(x)$ are the same, which in general does not hold true even if their spectra are the same.

\section{Explicit model}\label{MEASMOD}

Our example is about a Bose-Einstein condensate in a double-well trap, as described by the two-mode Hamiltonian~\cite{MIL97} or the two-site Hubbard model put in the form
\beq
H = -\tau J_x + \epsilon J_z + UJ_z^2\,.
\label{HAM}
\eeq
Given the boson annihilation operators for the left and right halves of the potential well $a_L$ and $a_R$, we have three operators that obey the angular momentum algebra,
\bea
J_x &=& \half(a^\dagger_L a_R+a^\dagger_R a_L),\label{JX}\\
J_y &=&\hbox{$\frac{1}{2i}$}(a^\dagger_L a_R-a^\dagger_R a_L),\label{JY}\\
J_z &=& \half(a^\dagger_L a_L-a^\dagger_R a_R)\label{JZ}\,.
\eea
The parameter $\tau$ is the tunneling amplitude for the atoms from one site to the other, $\epsilon$ is the energy difference for an atom in the two sites, and $U$ represents the strength of the atom-atom interactions. The relevant boson states are spanned by $\ket{n_L,n_R}$, where $n_L$ and $n_R$ are the numbers of the atoms in the left and right traps, or $\ket{Jm} = \ket{J+m,J-m}$ that are eigenstates of the angular momentum with the components~\eq{JX}-\eq{JZ} for the total angular momentum $J$ and its $z$ component $m$. For a fixed number $N=2J$ atoms, the state space has the dimension $N+1$. 

As has been known for quite a while~\cite{MIL97}, the validity of the two-mode model is no longer guaranteed when the atom-atom interaction per particle becomes comparable to the energy difference between the one-particle eigenstates in one or the other of the potential wells. This is the case with experiments in the limit of the Thomas-Fermi approximation. However, the atom-atom interaction parameter $U$ and the validity of the two-mode model may, in principle, be controlled independently by adjusting both the trapping frequency of the potential wells (say, by adjusting the intensity of the trap lasers) and the atom-atom scattering length (say, by making use of a Feshbach resonance). In what follows, the two-mode approximation is always assumed valid.

We take it that $\epsilon$ is the parameter to be determined. Time evolution according to the Hamiltonian~\eq{HAM} corresponds to the generator of measurement results according to $K=Ht/\hbar$, but we absorb the factor $t/\hbar$ into the definition of the parameters and make no difference between the operators $H$ and $K$.  The measurement is about a dimensionless evolution phase, with $\epsilon$ standing for what we denoted by $\theta$ in Sec.~\ref{GENMEAS}.

Such a measurement is to an extent a standard task. In the usual optical Mach-Zehnder interferometer $\epsilon$ would be proportional to the phase difference of light incurred between the two arms of the interferometer, and likewise in atom interferometer setups. If we momentarily ignore both the tunneling between the two sites and the interactions between the atoms, we have the usual situation for an atomic  or optical interferometer with
\beq
K = \epsilon J_z\,.
\eeq
Given $N$ bosons, a comparison with~\eq{SIMPLEMOD} and~\eq{SIMPLERES} shows that the maximum Fisher information and the corresponding input state are
\beq
F_M = N^2,\quad \ket{\psi_M} = \hbox{$\frac{1}{\sqrt 2}$}(\ket{N,0} + e^{i\varphi}\ket{0,N})\,.
\label{OPTIMUM}
\eeq
With $\varphi=0$, we manifestly have what is called the NOON state~\cite{LEE02}. It gives the best possible measurement precision with the standard deviation $\sigma_\epsilon = 1/\sqrt{F_M}=1/N$, the usual Heisenberg limit.

The assignment now is to find out how the best possible measurement precision is affected when tunneling and nonlinear boson-boson interactions are included. In fact, the added nonlinearity $\propto J_z^2$ that commutes with the generator $J_z$ in itself has no effect on the achievable measurement precision. The situation becomes nontrivial only when site-to-site tunneling proportional to the noncommuting operator $J_x$ is also involved.

Even though the parameters  $\tau$, $\epsilon$ and $U$ in the Hamiltonian have the dimension of energy and one of them could be arbitrarily picked as the unit of energy, because of the time evolution they get multiplied by $t/\hbar$ and are rendered dimensionless. This means that we already have three independent parameters. Moreover, on top of the double optimization over both the measurement process and the initial state as in Eqs.~\eq{FISHER} and~\eq{MFENT} we have another parameter to consider, the total number of bosons $N$. In that regard, in the special case with $\tau=0$ the maximum Fisher information is $N^2$. Second, given $N$ atoms, the effects of particle number in the Hamiltonian scale with various parameters approximately as $\tau N$, $\epsilon N$, and $U N^2$. To make the physics as invariant as possible with respect to the number of atoms, it pays to compare $\epsilon$ and $NU$. In short, we will express the results in terms of the scaled variables
\beq
f= F/N^2,\quad u = NU\,.
\eeq
The Heisenberg limit is given by $f=1$.

All numerical computations were done using Mathematica. To avoid the issue of degenerate versus nondegenerate perturbation theory, we advise to avoid parameter values precisely equal to zero. Even very small nonzero values that are in practice equal to zero eliminate the problems, presumably by reducing the symmetry of the Hamiltonian.

\section{Results}\label{MEASDEP}
\begin{figure}
\includegraphics[width=8.0cm]{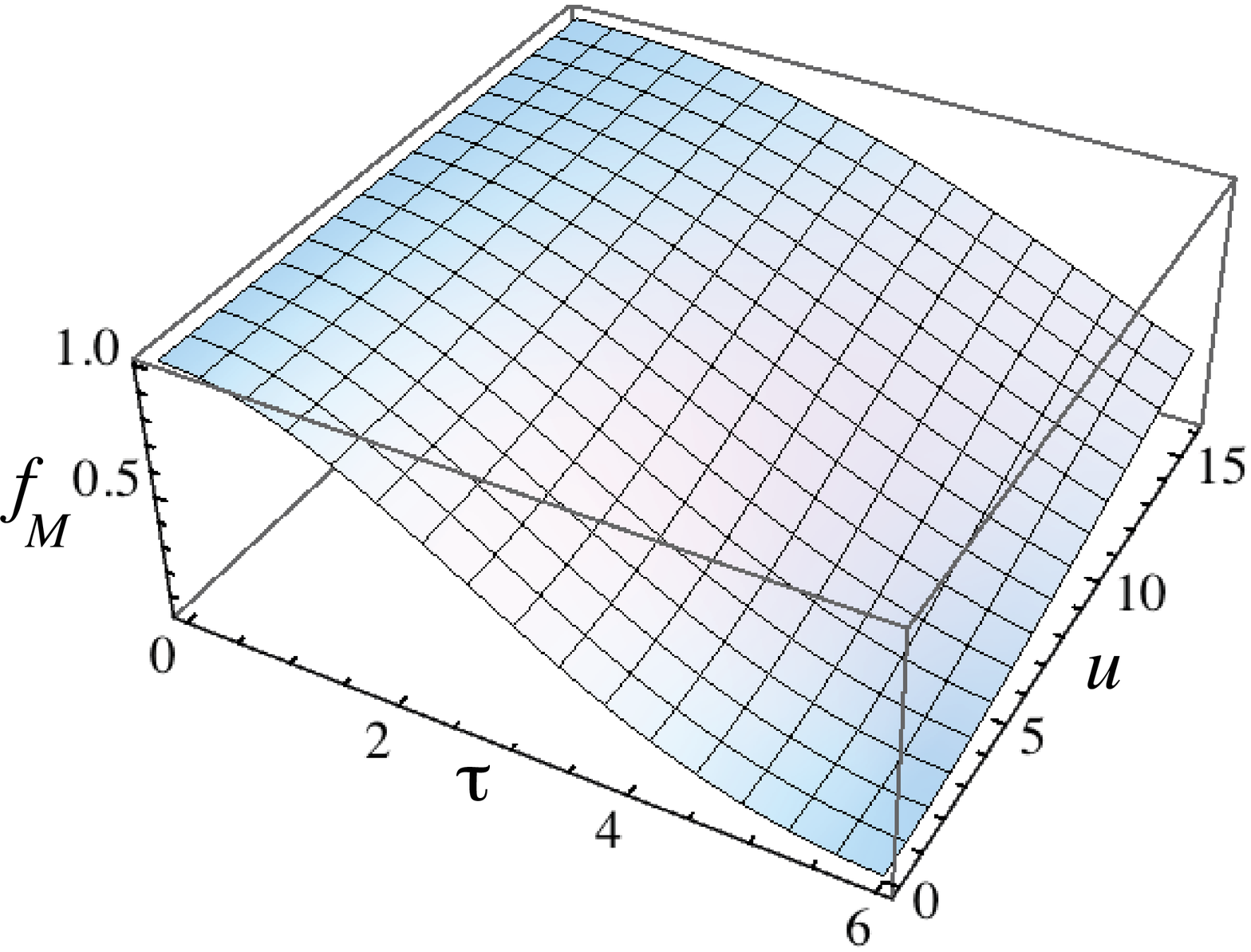}\\
\vspace{10pt}
\includegraphics[width=6.5cm]{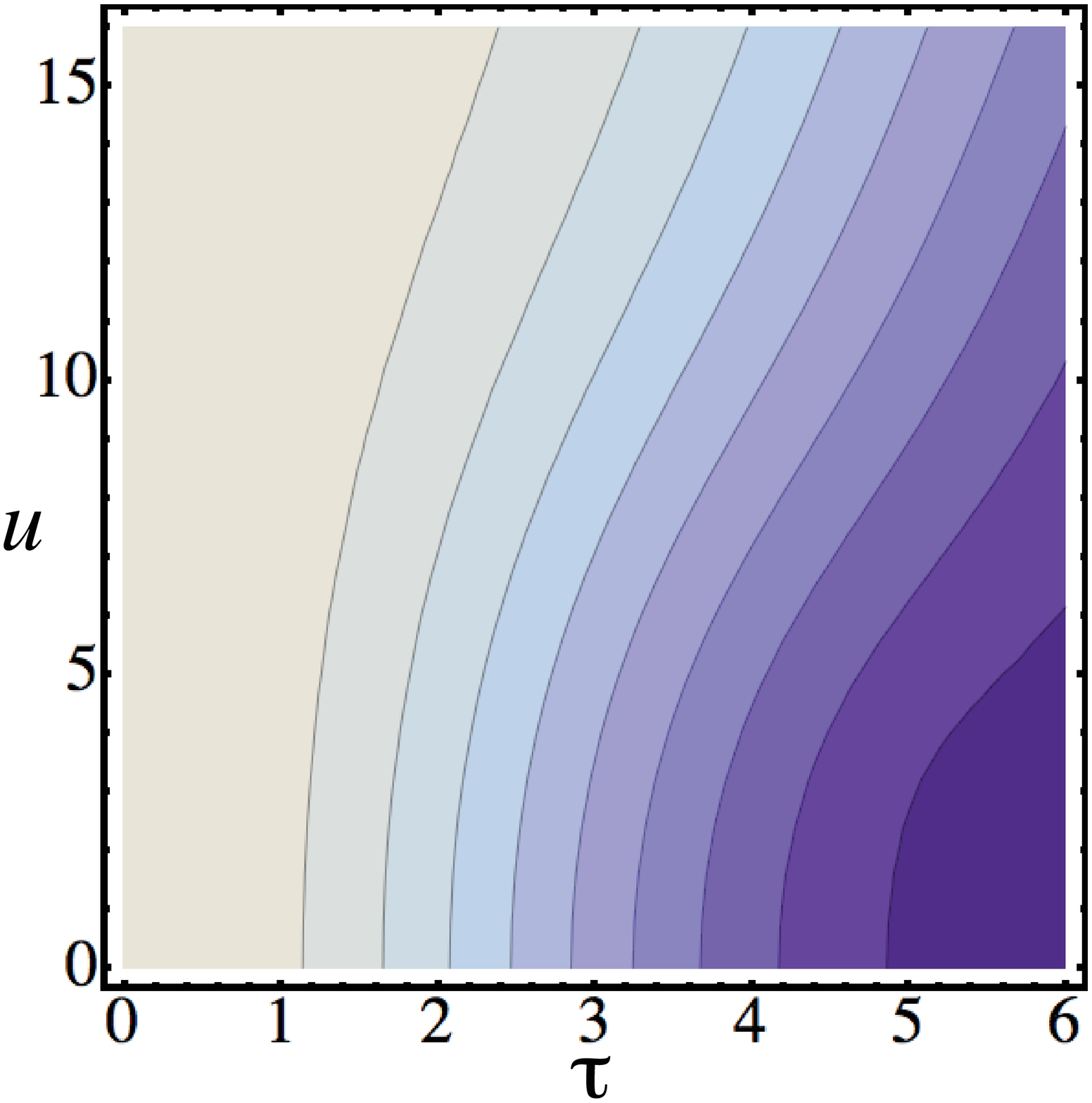}
\caption{(Color online) 3D and contour plots of the scaled maximal Fisher information $f_M=F_M/N^2$ as a function of the tunneling amplitude $\tau$ and scaled atom-atom interaction $u=NU$, given the atom number $N=2$ and energy difference $\epsilon=1$.}
\label{TWOAT}
\end{figure}

Let us first ignore atom-atom interactions and write
\bea
H &=&T(\sin\phi\, J_x + \cos\phi\, J_z);\label{NONON}\\
 T &=& \sqrt{\tau^2+\epsilon^2},\, \cos\phi=\epsilon/T,\, \sin\phi=-\tau/T\,.
\eea
The Hamiltonian~\eq{NONON} is proportional to the component of the  (fictitious) angular momentum in the direction $\cos\phi\, \hat{\bf e}_z+ \sin\phi\, \hat{\bf e}_x$ leaning an angle $\phi$ away from the $z$ axis. The corresponding angular momentum eigenstates $\ket{Jm}_\phi$ are eigenstates of the Hamiltonian, and the eigenvalues are $\lambda_m = mT$.

Computation of the local generator $L$ from Eq.~\eq{EXPLFRM} is a tedious affair involving the matrices that govern the transformations of the eigenstates of angular momentum under rotations~\cite{EDM60}, but we may produce a simple analytical example in the limit $T\rightarrow\infty$. Then the eigenvalues of the Hamiltonian $\lambda_m$ tend to infinity, in Eq.~\eq{EXPLFRM} only the diagonal elements with $m=n$ survive, and we have the matrix elements of the generator of local translations
\beq
{}_\phi\!\mele{Jm}{L}{Jm'}\!{}_\phi = \delta_{mm'}\, {}_\phi\!\mele{Jm}{J_z}{Jm'}\!{}_\phi\,.
\eeq

The expectation value of a component of an angular momentum in an eigenstate of the component of the angular momentum tilted by an angle $\phi$ is in fact governed by the classical projection, so we have the diagonal elements of the matrix $L$, and at the same time its eigenvalues $\ell_m$, in the form $\ell_m = m \cos\phi$. The maximal Fisher information is therefore $F_M = J^2/4 \,\cos^2\phi=N^2\,\cos^2\phi$, and the measurement uncertainty increases by a factor of $1/|\cos\phi|$ compared to the case without tunneling. In particular, for small values of the energy difference $\epsilon$ the tunneling reduces the measurement precision by a factor of $\simeq|\epsilon/\tau|$.

We next re-instate atom-atom interactions. We begin in Fig.~\ref{TWOAT} with the  case $N=2$. Panel 1(a) is a 3D plot of the scaled Fisher information $f_M$ maximized over all input states as a function of the tunneling and atom-atom interaction parameters $\tau$ and $u$, with $\epsilon=1$. We have plotted only nonnegative values of $u$, meaning, repulsive atom-atom interactions. The Fisher information is not exactly an even function of $u$, but the difference between $u$ and $-u$ for the purposes of these drawings is so small that it would be barely discernible in Fig.~\ref{TWOAT}, or in Fig.~\ref{LIMIT} below. As expected, we have the Heisenberg limit all along the $\tau=0$ axis. Increasing the tunneling amplitude $\tau$ decreases the maximum Fisher information, while increasing atom-atom interactions counteracts the effect of tunneling~\cite{GRO11}.

As a matter of fact, in the formal limit $|u|\rightarrow\infty$ the eigenstates of the Hamiltonian approach the eigenstates of $J^2_z$. We will eventually have doubly degenerate manifolds made of states $\ket{Jm}$ and $\ket{J-\!\!m}$, except for $m=0$ when there is no degeneracy. Perturbation theory requires that we diagonalize $J_z$ within each of these manifolds, whereupon we need to pick the eigenstates of the Hamiltonian to be eigenstates of $J_z$ itself. Therefore, in the case $|u|\rightarrow\infty$ Eq.~\eq{EXPLFRM} shows that the local generator equals the original generator, $L=J_z$, and the Heisenberg limit ensues.

Although we generally say little about the optimal input state, we mention here the  special case $u\rightarrow -\infty$. It is known~\cite{CIR98} that in this limit, and with $\epsilon=0$, the zero-temperature ground state is a ``Schr\"odinger cat'' or NOON state of the form~\eq{OPTIMUM} that maximizes the Fisher information. The practical complications are probably substantial, but in principle the right input state comes almost for free.

\begin{figure}
\begin{center}
\includegraphics[width=8cm]{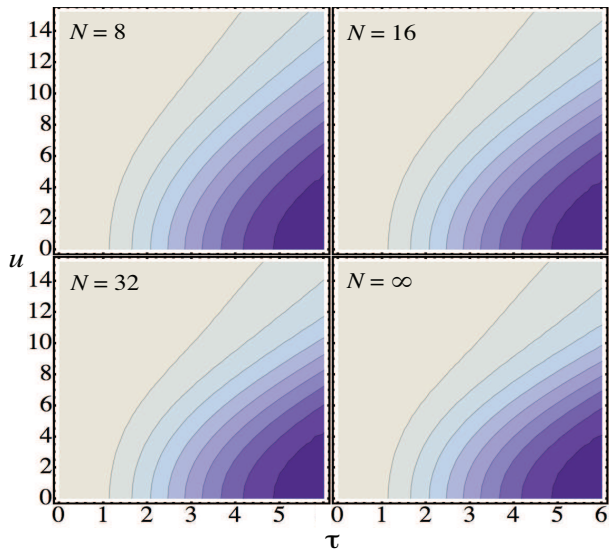}
\end{center}
\caption{(Color online)  Contour plots of $f_M$ as a function of $\tau$ and $u$ with a fixed $\epsilon=1$ for $N=8$, 16 and 32, as marked on the panels. Also shown is an extrapolation to the limit $N\rightarrow\infty$ denoted by $N=\infty$. The maximum value of $f_M$ is attained at $\tau=0$, and equals $f_M=1$. The spacing between the contour lines equals 0.1.}
\label{LIMIT}
\end{figure}

Figure~\ref{LIMIT} gives similar contour plots as Fig.~\ref{TWOAT}, but for different atom numbers. We firstly have $N=8$, 16, and 32. The contours are all at the same values of $f_M$. It is obvious that in the limit $N\rightarrow\infty$ the scaled maximum Fisher information $f_M=F/N^2$ as a function of the parameters $\tau$ and $u=NU$ converges to a universal function $f_M(\tau,u,\epsilon,N\rightarrow\infty)$.  While the Heisenberg scaling of measurement uncertainty is distinctly a quantum effect, this is also  the limit in which one expects that the semiclassical approximation becomes valid: Barring pathological quantum states of the system such as the Schr\"odinger cat (sic!), $b_L$ and $b_R$ may be treated as classical variables with certain Poisson brackets  instead of quantum operators.

We have investigated the convergence with increasing $N$  for a few fixed values of $\tau$ and $u$. It appears that for large $N$ the deviation of $f_M$ from the $N\rightarrow\infty$ limit scales as $N^{-1}$. Based on this observation, we have done a Richardson extrapolation~\cite{NUMRES} to the limit function $f_M(\tau,u,\epsilon,N\rightarrow\infty)$, and plot it  in Fig.~\ref{LIMIT} labeled as $N=\infty$. The absolute error of the limit function is conservatively estimated to be less than $10^{-3}$ in the whole plotted range, which is far below the resolution of the plot. The limit function is similar to the $N=2$ function $f_M(\tau,u,\epsilon,2)$, and the preceding qualitative discussions of this case still apply.

So far we have set the value of the quantity to be measured as $\epsilon=1$; next we address the variation of Fisher information with the parameters $\epsilon$ itself. One conceivable reason is that the parameters $\epsilon$, $\tau$ and $u$ all scale simultaneously with the interaction time. Accordingly, for the time being we write $\epsilon\rightarrow x\epsilon$ with a scaling factor $x$, and similarly for $\tau$, $u$. 

In the limit $x\rightarrow0$ one finds that $K\rightarrow0$, $e^{-iK}\rightarrow 1-iK$, $\ket{\psi'}\rightarrow -iJ_z \ket{\psi}$, and we are back to the Heisenberg limit. However, the measured value of $\epsilon$ scales to zero as well, and the relative accuracy becomes poor. 

To address the opposite limit $x\rightarrow\infty$ we note that $K$ and its eigenvalues $\lambda_n$ grow linearly proportional to the scaling factor $x$. In the limit of a large $x$ only the diagonal term with $m=n$ survives in the sum in Eq.~\eq{EXPLFRM}, and we have a local generator $L$ that only retains the part of $J_z$ diagonal in the eigenbasis of $K$,
\beq
L \simeq \sum_n \ket{n}\mele{n}{J_z}{n}\bra{n}\,.
\eeq
Numerically, we correspondingly see that the maximum precision converges to a constant,  and with $N\rightarrow\infty$ apparently to a constant fraction of the Heisenberg limit.

The remaining limits we consider are for varying $\epsilon$ with the other parameters held constant. In the case $\epsilon\rightarrow\infty$ the other terms in the Hamiltonian become insignificant perturbations, and the Heisenberg limit as for $K = \epsilon J_z$ is reached. In the contrary limit $\epsilon\rightarrow0$ the measurement precision tends to a nonzero constant. The way it works in the lowest order in $\epsilon$ is seen from Eq.~\eq{EXPLFRM}: Calculate the vectors $\ket n$ and eigenvalues $\lambda_n$ simply by setting $\epsilon=0$ in the operator $K$, find the local generator $L$, and obtain the maximal Fisher information from the range of its eigenvalues.  This limit is of some interest as detection and measurements of small energy differences $\epsilon$ is conceivably a frequent task. It is, however, not worth the space to draw a new figure,  as the case $\epsilon\simeq0$ is effectively realized whenever $\epsilon\le J$ and $\epsilon\le |u|$. If plotted  with the same axes as in Fig.~\ref{LIMIT}, the $\epsilon=0$ results would be virtually indistinguishable from the $\epsilon=1$ results already shown.

\section{Concluding remarks}\label{CONCLUSIONS}

It should be understood that we approach the Heisenberg limit purely from the perspective of principles. Except for one incidental remark we pay no heed to the question of preparing the optimal probe state, nor to practical measurement strategies, nor to the problems with imperfect real experiments. With these caveats, we have studied the combined effects of nonlinearity and ``arm-to-arm'' tunneling on the best possible measurement precision of an interferometer. We use measurements of the difference of the energies of the potential wells in a double-well trap for bosonic atoms as a tangible example. We have adapted time independent perturbation theory to prove that the Heisenberg limit cannot be beat indirectly as a result of the nonlinearity due to atom-atom interactions, and to study numerically the behavior of the measurement precision with varying problem parameters. Many of the limiting cases are also easy to understand on the basis of the perturbation theory. As an interesting aside, we have noted that in the limit of very strong attractive atom-atom interactions the input state required for Heisenberg limit precision is in fact a zero-temperature ground state. 

\section{Acknowledgments}
This work is supported in part by NSF, Grant No. PHY-0967644.
\end{document}